\documentclass[12pt]{iopart}

\usepackage{epsfig}

\begin{document}

\title{ Kalb-Ramond scalar $QED$ multiple vacua }

\author{Anais Smailagic$^1$, Euro Spallucci$^2$}

\address{$^1$ INFN, Sezione di Trieste, Trieste, Italy}
\ead{anais@ts.infn.it }
\address{ $^2$INFN, Sezione di Trieste, Trieste, Italy}
\ead{Euro.Spallucci@ts.infn.it}

\begin{abstract}
We study a model of interacting vector and Kalb-Ramond  gauge fields in a non-trivial Higgs vacuum generated by a charged 
and a neutral scalar field.  The system admits different vacua for different v.e.v. of the two Higgs fields. 
Our primary interest in this paper regards the "~\emph{mixed phase}~" where both the photon and the Kalb-Ramond acquire a mass. 
In this phase we compute the interaction potential energy between static test charges. It turns out that the limit in which 
the photon becomes massless, while the Kalb-Ramond remains massive, leads to a Cornell confining potential between test charges.    
 \end{abstract}



\maketitle

\section{Introduction}
 
The distinctive feature of Strong Interactions is that elementary constituents (quarks and gluons) are permanently
confined inside hadrons. This idea  is supported by the  empirical evidence that no high energy experiment results in the
detection of free quarks or gluons. It was only possible, so far, to have an indirect evidence of confinement through 
the production of hadronic jets.\\
The theoretical description of Strong Interaction in terms of  Yang-Mills gauge theory, is able to describe
well only the asymptotically free regime at short distances, where quarks and gluons behave as almost free particles. 
On the other hand, no satisfactory derivation of the confining potential at large distances has been obtained in this 
theoretical framework, since in the strong-coupling regime perturbation theory cannot be applied. 
However, several papers have put forward the idea that the confining effect should be attributed to the Abelian sub-group 
of the $SU\left(\,N\,\right)$ symmetry group of $QCD$. The original hypothesis of \emph{Abelian domiance} \cite{Ezawa:1982bf,Ezawa:1982ey},
has been investigated in depth by many authors, e.g.
\cite{Luscher:1978rn,tHooft:1981bkw,Suzuki:1988qa,Maedan:1988yi,Kondo:1997pc,Kondo:1997kn,Deguchi:2002cr,Deguchi:2003zi,Kondo:2014sta}.\\
Until this matter awaits a solution at the theoretical level, the spectrum of heavy mesons can be described
in terms of ``~cooked-up~'' phenomenological potentials that fit well the experimental data. 
The most successful of these  attempts is the so-called Cornell potential which is simply 
a Coulomb potential with a linear tail \cite{Eichten:1978tg}.\\
In a recent paper  we addressed this problem in the  framework of an Abelian gauge theory  in which a 
$U\left(\,1\,\right)$ vector field  was coupled to a Kalb-Ramond tensor field \cite{Smailagic:2020kep} .
The important role of maximal rank gauge field strength in relation with confinement and scale symmetry
breaking has been discussed in several papers 
\cite{Ansoldi:2002id,Guendelman:2003ib,Gaete:2006xd,Gaete:2006nq,Gaete:2007ry}
Furthermore, a suitable  Higgs mechanism was designed in order to give
mass to one of the two fields respectively. In the vacuum of a massive Kalb-Ramond field we were able to 
obtain the Cornell potential analytically.\\
In this paper, we continue along these lines and investigate a more general scenario in which both fields can acquire 
mass simultaneously and independently of each other.
The possibility to make different choices in the Higgs potential will lead to a variety of possible vacua. 
For each vacuum we shall compute
the corresponding interaction potential energy between two static test charges.  We expect that the form of the 
 resulting interaction potential will crucially depend on the values of the mass parameters in the Higgs potential.
\\ 
In Section(\ref{lwed}) we describe the choice of the Higgs mechanism giving rise to  masses of both photon and Kalb-Ramond 
fields. Since we are interested  in an \emph{effective} theory for the vector field in the non-trivial Higgs vacuum
 we shall first integrate
out the Goldstone mode and successively the Kalb-Ramond field. Thus, we end up with a non-local and
$U\left(\,1\,\right)$  gauge invariant Lagrangian, in spite of the presence of the mass parameters. From this effective
Lagrangian we obtain the expression for the interaction energy between two charges in the static limit.\\
In Section(\ref{leggero}) we explicitly calculate the interaction energy in two different phases corresponding to
different ratios of  mass parameters.  Finally, we discuss the transition between those phases.\\
In the last Section(\ref{ciacole}) we discuss the results and comment on the possible implications to the phenomenological
interesting non-Abelian case. 

\section{Kalb-Ramond $QED$}
\label{lwed}

As mentioned in the Introduction, we write down an  Abelian gauge field  $A_\mu$ coupled to a Kalb-Ramond $2$-form $B_{\mu\nu}$. 
The two fields interact with a charged  and neutral Higgs fields respectively. The Lagrangian density for this model
 is given by  

\begin{eqnarray}
&&\mathcal{L}\left[\, A\ ,B\ ,\phi\ ,\phi^\ast\ ,\psi\,\right] =
\frac{1}{4}\,F_{\mu\nu}^2\left(\,A\,\right)+\frac{1}{2\times 3!}\,H_{\mu\nu\rho}^2\left(\,B\,\right)
+\left(\, D_\mu\left(\,A\,\right)\phi\,\right)^\ast\, D^\mu\left(\,A\,\right)\phi \nonumber\\
&&+\frac{1}{2}\partial_\mu\psi\, \partial^\mu\psi -V\left(\,\phi\ ,\phi^\ast\ ,\psi\,\right)
+\frac{g^2}{4} \psi^2\, B_{\mu\nu}^2 -\frac{g}{2}\psi\,B^\ast_{\mu\nu}\, F^{\mu\nu}\left(\,A\,\right) -e \, J^\mu_{ext} \, A_\mu
\label{1}
\end{eqnarray}

where the $U\left(\,1\,\right)$ covariant derivative and the Higgs potential read
\begin{eqnarray}
&& D_\mu\left(\,A\,\right)\phi\equiv \left(\, \partial_\mu -ie\,A_\mu\,\right)\,\phi\ ,\\
&& V\left(\,\phi\ ,\phi^\ast\ ,\psi\,\right)\equiv -\frac{1}{2}\mu^2_\phi \phi^\ast\phi -\frac{1}{2}\mu^2_\psi \,\psi^2 +\frac{\lambda_\phi}{6}\left(\, \phi^\ast\,\phi\,\right)^2+\frac{\lambda_\psi}{4!}\,\psi^4
\end{eqnarray}

One notices the presence of two independent mass parameters  $\mu_\phi$ and $\mu_\psi$ associated to the charged and 
to the neutral Higgs field respectively.\\
Being interested in the evaluation of the interaction energy between two static, point-like, test charges  we also introduced
an \emph{external} current in (\ref{1}). \\
For convenience let us choose the Euler decomposition for the complex field $\phi$:

\begin{equation}
\phi\left(\, x\,\right)\equiv \frac{1}{\sqrt{2}}\rho\left(\, x\,\right)\, e^{i\theta\left(\, x\,\right)}
\end{equation}

and  rewrite (\ref{1}) as

\begin{eqnarray}
\mathcal{L}\left[\, A\ ,B\ ,\rho\ ,\theta\ ,\psi\,\right] &&=
\frac{1}{4}\,F_{\mu\nu}^2\left(\,A\,\right)
+ \frac{1}{2}\left[\,\partial_\mu\rho\,\partial^\mu\rho  + \rho^2\,\left(\,\partial_\mu\theta\,\right)^2 
-2e\rho^2 A^\mu \partial_\mu\theta +e^2\,\rho^2 \,A^2 \,\right]\nonumber\\
&&+\frac{1}{2}\partial_\mu\psi\, \partial^\mu\psi -V\left(\,\rho\ ,\psi\,\right)
+\frac{1}{2\times 3!}\,H_{\mu\nu\rho}^2\left(\,B\,\right)\nonumber\\
&&+\frac{g^2}{4} \psi^2\, B_{\mu\nu}^2 -\frac{g}{2}\,\psi\,B^\ast_{\mu\nu}\, F^{\mu\nu}\left(\,A\,\right) -e \, J^\mu_{ext} \, A_\mu
\label{5}
\end{eqnarray}

Due to the $U(1)$ symmetry the Higgs potential is independent from the Goldstone boson $\theta$
and has \emph{stationary} points given by 

\begin{eqnarray}
&&  \frac{\partial V}{\partial \rho}=-   \mu^2_\phi\, \rho  +\frac{\lambda_\phi}{6}\, \rho^3=0   \ ,\\
&&\frac{\partial V}{\partial \psi}=-   \mu^2_\psi\, \psi  +\frac{\lambda_\psi}{6} \,\psi^3=0
\end{eqnarray}

The above equations give four different ``~vacua~''

\begin{eqnarray}
&& (A)\quad : \quad \rho_0=0\ ,\qquad \psi_0=0\ ,\qquad\hbox{Coulomb vacuum}\ ,\\
&& (B)\quad : \quad \rho_0^2=\frac{6\mu^2_\phi}{\lambda_\phi}\ ,\qquad \psi_0=0\ ,\qquad\hbox{Yukawa vacuum}\ ,\\
&& (C) \quad : \quad \rho_0=0\ ,\qquad \psi_0^2=\frac{6\mu^2_\psi}{\lambda_\psi}\ ,\qquad\hbox{Cornell vacuum }\ ,\\
&& (D)\quad : \quad \rho_0^2=\frac{6\mu^2_\phi}{\lambda_\phi}\ ,\qquad 
\psi_0^2=\frac{6\mu^2_\psi}{\lambda_\psi}\qquad\hbox{Mixed vacuum}
\end{eqnarray}

The first three vacua have been  studied in detail in a previous work \cite{Smailagic:2020kep}. 
A similar phase portrait has been recently obtained 
in a recent work  where, instead of a vector and a tensor, two different vector fields have been coupled
to a pair of Higgs fields \cite{Scott:2018xgo}.
In this paper we shall concentrate
on the properties of the mixed phase $(D)$ as it interpolate between the limiting cases $(B)$ and $(C)$.
In the mixed vacuum the Lagrangian (\ref{5}) reads

\begin{eqnarray}
&&\mathcal{L}\left[\, A\ ,B\ ,\theta\,\right] =
+\frac{1}{4}\,F_{\mu\nu}^2\left(\,A\,\right)+\frac{1}{2\times 3!}\,H_{\mu\nu\rho}^2\left(\,B\,\right)+\nonumber\\
&&+\frac{1}{2}\left[\, \rho^2_0\,\left(\,\partial_\mu\theta\,\right)^2 
-2e\,\rho^2_0\, A^\mu \,\partial_\mu\theta+e^2\,\rho^2_0 \,A^2\,\right]+\nonumber\\
 &&-V\left(\,\rho_0\ ,\psi_0\,\right)+\frac{g^2}{4} \psi^2_0\, B_{\mu\nu}^2 
-\frac{g}{2} \psi_0\,B^\ast_{\mu\nu}\, F^{\mu\nu}\left(\,A\,\right) -e \, J^\mu_{ext} \, A_\mu
\label{l1}
\end{eqnarray}
We can drop the constant term $V\left(\,\rho_0\ ,\psi_0\,\right)$, being irrelevant, since no gravitational effects 
are considered  in this model.\\
The presence of the Goldstone boson persists in (\ref{l1}) but can be eliminated using its field equation

\begin{equation}
\partial^2\theta= e \partial A\longrightarrow \theta= e\,\frac{1}{\partial^2} \partial A
\label{gold}
\end{equation}

Lagrangian (\ref{l1}) becomes 

\begin{eqnarray}
&&\mathcal{L}\left[\, A\ ,B\,\right] =
\frac{1}{4}\,F_{\mu\nu}\left[\, 1 -\frac{e^2\rho_0^2}{\partial^2}\,\right]\,F^{\mu\nu}-e \, J^\mu_{ext} \, A_\mu  \nonumber\\
 &&+\frac{1}{2\times 3!}\,H_{\mu\nu\rho}^2\left(\,B\,\right)+\frac{g^2}{4} \psi^2_0\, B_{\mu\nu}^2 -\frac{g}{2}\psi_0\,\widetilde{B}_{\mu\nu}\, F^{\mu\nu}\left(\,A\,\right) 
\label{l2}
\end{eqnarray}

The next step is to eliminate the Kalb-Ramond field in favor of $A_\mu$ by using the field equation

\begin{equation}
B_{\mu\nu}=-\frac{g\psi_0}{\partial^2 -g^2\psi^2_0} \,F^\ast_{\mu\nu} \label{17}
\end{equation}

Inserting (\ref{17}) in (\ref{l2}) one finds an \emph{effective} Lagrangian for $A_\mu$ as

\begin{equation}
\mathcal{L}\left[\, A\,\right] =
+\frac{1}{4}\,F_{\mu\nu}\left[\, 1 -\frac{e^2\rho_0^2}{\partial^2}+\frac{g^2\psi_0^2}{\partial^2 -g^2\psi_0^2}\,\right]
\,F^{\mu\nu}-e \, J^\mu_{ext} \, A_\mu  
 \label{l3}
\end{equation}

The Lagrangian (\ref{l3}) for $A_\mu$ describes a non-local version of electrodynamics, while still \emph{preserving gauge
symmetry}, in spite of the presence of mass parameters. This interesting result is due to the fact that we integrated out
the Goldstone boson instead of choosing a particular value of $\theta$ (gauge fixing). 
The resulting modification of the standard electrodynamics can be seen as
corresponding to standard electromagnetic interaction but in a ``~medium~'' with a non-trivial dielectric constant 
depending on $\rho_0$ and $\psi_0$ \cite{Smailagic:2020bgf}.\\
The v.e.v.'s  labeling   phases $(A)$, $(B)$, $(C)$ and the corresponding potentials are:

\begin{itemize}
\item  $\rho_0=0$, $\psi_0=0$ gives the usual Coulomb electrodynamics
\begin{equation}
\mathcal{L}\left[\, A\,\right] =
\frac{1}{4}\,F_{\mu\nu}\,F^{\mu\nu}-e \, J^\mu_{ext} \, A_\mu  
\end{equation}

\begin{equation}
V_{int}\left(\, r\,\right)= -\frac{e^2}{4\pi}\frac{1}{r} 
\end{equation}

\item $\rho_0^2 = 6\mu^2_\phi/\lambda_\phi$, $\psi_0=0$ gives the Yukawa short-range interaction 

\begin{equation}
\mathcal{L}\left[\, A\,\right] =
\frac{1}{4}\,F_{\mu\nu}\left[\, 1 -\frac{e^2\,\rho_0^2}{\partial^2} \,\right]\,F^{\mu\nu}-e \, J^\mu_{ext} \, A_\mu  
 \label{l4}
\end{equation}

\begin{eqnarray}
V_{int}\left(\, r\,\right)=&& -\frac{e^2}{4\pi}\frac{1}{r}\, e^{-e\rho_0\,r}
\end{eqnarray}

\item $\rho_0=0$, $\psi_0^2=6\mu^2_\phi/\lambda_\psi$ describes the ``~\emph{charge confinement}~'' characterized 
by a Cornell potential
\begin{equation}
\mathcal{L}\left[\, A\,\right] =
+\frac{1}{4}\,F_{\mu\nu}\left[\, \frac{\partial^2}{\partial^2 -g^2\psi^2_0}\,\right]\,F^{\mu\nu}-e \, J^\mu_{ext} \, A_\mu  
 \label{lcornell}
\end{equation}
\begin{equation}
V_{int}\left(\, r\,\right)= -\frac{e^2}{4\pi}\,\frac{1}{r} + \frac{e^2g^2 \psi_0^2}{8\pi}\,r
\end{equation}
\end{itemize}
The same result, but with a repulsive Coulomb interaction, has been obtain in 
\cite{Chatterjee:2016liu,Mukherjee:2019vmi} to model Cooper pairs in a super conducting
medium.\\

In this paper we are interested in the study of the 
\emph{``~Mixed Phase~''} $(D)$. \\
From (\ref{l3}) one  solves the field equations for $F_{\mu\nu}$ in terms of $J^\mu_{ext}$ and obtains

\begin{equation}
F^{\mu\nu}
= e \,\partial^{[\, \mu}\frac{\partial^2 -g^2\psi^2_0}{\left(\,\partial^2\,\right)^2 -e^2\rho_0^2\,\partial^2 
+e^2 g^2\rho_0^2 \psi_0^2}\,J^{\nu\,]}_{ext}\label{effe}
\end{equation}

Inserting (\ref{effe}) in (\ref{l3})  one finds the effective Lagrangian

\begin{eqnarray}
 L\left[\,J\right]
&&=\frac{e^2}{2} \, J^\mu_{ext} \,\frac{\partial^2 -g^2\psi_0^2}
{\left(\,\partial^2\,\right)^2 -e^2\,\rho_0^2\,\partial^2 +e^2g^2\rho_0^2 \psi_0^2}\,J_{ext\,\mu}\ ,\nonumber\\
&&=\frac{e^2}{2} \, \int d^4y J^\mu_{ext}\left(\,y\,\right)\, G\left(\, y -x\,\right) \,J_{\mu\, ext}\left(\,x\,\right)
\label{lj}
\end{eqnarray}

where we introduced the Green function $ G\left(\, y -x\,\right)$ defined by 
\begin{equation}
 \frac{\left(\,\partial^2\,\right)^2 -e^2\rho_0^2\,\partial^2 
+e^2 g^2\rho_0^2 \psi_0^2}{\partial^2 -g^2\psi_0^2}G\left(\, y -x\,\right)=\delta\left(\, x-y\,\right)
\end{equation}
From (\ref{lj}) one obtain the effective action which is useful to obtain the interaction energy

\begin{equation}
 W\left[\, J \,\right] \equiv \int d^4x L\left[\,J\right]= \int d^4x \int d^4 
J^\mu_{ext}\left(\,y\,\right)\, G\left(\, y -x\,\right) \,J_{\mu\, ext}\left(\,x\,\right)
\label{wj}
\end{equation}

  We choose $J^\mu_{ext}$ to represent  two heavy \emph{static charges} sitting in $\vec{x}_1$ and $\vec{x}_2$.

\begin{equation}
J^\mu_{ext}\left(\, x \,\right) = \delta^\mu_0 \left[\, \delta^{(3)}\left(\, \vec{x}-\vec{x}_1\,\right)-
\delta^{(3)}\left(\, \vec{x}-\vec{x}_2\,\right)\,\right]
\end{equation}

The cross terms in (\ref{wj}) defines the mutual interaction energy 
$V\left(\,\vec{x}_1\ , \vec{x}_2\,\right)$  as

\begin{eqnarray}
V_{int}\left(\,\vec{x}_1\ , \vec{x}_2\,\right)&&= e^2 \int d^3x\,
 \delta^{(3)}\left(\, \vec{x}-\vec{x}_1\,\right)\times\nonumber\\
&&\frac{\nabla^2_x -g^2\psi_0^2}
{\left(\,\nabla^2_x \,\right)^2 -e^2\,\rho_0^2\,\nabla_x^2 +e^2g^2\rho_0^2 \psi_0^2}
\delta^{(3)}\left(\, \vec{x}-\vec{x}_2\,\right)\label{28}
\end{eqnarray}

We omit the self-interaction energy for each charge.\\
In order to calculate the explicit form of (\ref{28}) we perform a 
Fourier transform of the delta-functions and obtain

\begin{equation}
 V_{int}\left(\,\vec{r}\,\right)= e^2 \int \frac{d^3k}{\left(\, 2\pi\,\right)^3} 
\frac{\vec{k}{}^2 + g^2\psi_0^2}{\vec{k}{}^4 +e^2 \,\rho_0^2\,\vec{k}{}^2 +e^2g^2\rho_0^2 \psi_0^2} e^{i\vec{k}\cdot\vec{r}}\ ,\quad
\vec{r}\equiv  \vec{x}_1-\vec{x}_2 \label{vint}
\end{equation}
For notational convenience, we introduce the photon ($\mu_A$) and Kalb-Ramond ($\mu_B$) mass   parameters as follows

\begin{eqnarray}
 && \mu^2_A\equiv e^2 \rho_0^2\ ,\\
 && \mu^2_B\equiv g^2 \psi_0^2
\end{eqnarray}

To calculate the  Fourier integral in (\ref{vint}) we find the poles of the integrand function determined 
by the quartic equation

\begin{equation}
 \vec{k}{}^4 + \mu^2_A\,\vec{k}{}^2 + \mu^2_A\mu^2_B=0
\end{equation}

The zeros are given by

\begin{equation}
 \vec{k}{}^2_\pm = -m^2_\pm=\frac{\mu^2_A}{2}\left[\,1 \mp \sqrt{1- \frac{4\mu^2_B}{\mu^2_A}}\,\right]
\label{poli}
\end{equation}

where $m^2_\pm$ can be complex conjugate, or real, depending whether $\mu_B >\mu_A/2$, or $\mu_B <\mu_A/2$.\\
Our primary interested is to recover in the proper  limit ($ \mu_A\to 0$) the Cornell potential. This limit is accessible
only in case when $\mu_A <2\mu_B$ which we call the ``~light photon phase~'' and will describe in detail in the next Section.

\section{``Light photon'' phase}
\label{leggero}
In the ``~light photon~'' phase  the Feynmann propagator  has complex poles

\begin{equation}
  \mu^2_A < 4\mu^2_B\ , \quad m^2_\pm=\frac{\mu^2_A}{2}\left(\, 1 \pm i\gamma \,\right)\ ,\quad
\gamma \equiv\sqrt{\frac{4\mu^2_B}{\mu^2_A}-1}
\end{equation}

which implies that $m_\pm^2$ are complex conjugate. For an easier handling, the Feynman propagator quartic in momenta 
is expressed  in terms of ordinary propagators as 

\begin{equation}
 \frac{\vec{k}{}^2 + \mu^2_B}{\vec{k}{}^4 +\mu^2_A\,\vec{k}{}^2 +\mu^2_A\mu^2_B}=
\frac{\vec{k}{}^2 + \mu^2_B}{m^2_- - m^2_+}\left[\, \frac{1}{\vec{k}{}^2 + m^2_+  } - 
\frac{1}{\vec{k}{}^2 + m^2_- }\,\right]\label{deco}
\end{equation}

Using the decomposition (\ref{deco}), the interaction potential (\ref{vint}) becomes

\begin{eqnarray}
&& V_{int}\left(\, r\,\right)=V_+\left(\, r\,\right)-V_-\left(\, r\,\right)\ ,\\
&& V_\pm\left(\, r\,\right)=-\frac{e^2}{\Delta m^2} \,\int\frac{d^3k}{\left(\, 2\pi\,\right)^3} e^{i\vec{k}\cdot \vec{r}}
 \int_0^\infty ds\,\left(\,\vec{k}{}^2 + \mu^2_B\,\right) \, e^{-s\left(\,\vec{k}{}^2 +m^2_\pm \,\right)}\ ,\\
&& \Delta m^2\equiv m^2_- - m^2_+ = -i\mu^2_A \, \gamma
\end{eqnarray}

where we used  the Schwinger representation of the propagators. Integrating out the momenta leads to

\begin{equation}
 V_\pm\left(\, r\,\right)=-\frac{e^2}{8\pi^{3/2}}\frac{1}{\Delta m^2} 
 \,\left(\, m_\pm^2 -\mu^2_B\,\right)
\,\int_0^\infty \frac{ds}{s^{3/2}}\, e^{-s\, \frac{\mu^2_A}{2} -\frac{r^2}{4s}}\,
e^{\pm is\, \frac{\mu^2_A}{2}\gamma}
\end{equation}

The  above integrals can be exactly  calculated and give the interaction potential energy of the form

\begin{eqnarray}
 V_{int}\left(\, r\,\right) && =\frac{e^2}{4\pi\, r}\frac{1}{i\mu_A^2\gamma} 
 \,\left[\, 
 \left(\, \frac{\mu^2_A}{2}-\mu^2_B\,\right)\left(\, e^{-\frac{\mu_A r}{\sqrt{2}} \sqrt{1-i\gamma} } -
 e^{-\frac{\mu_A r}{\sqrt{2}} \sqrt{1+i\gamma} } \,\right)\right.\nonumber\\
&&  \left.  -i\frac{\mu^2_A \gamma}{2}
\left(\, e^{-\frac{\mu_A r}{\sqrt{2}} \sqrt{1-i\gamma} } +
 e^{-\frac{\mu_A r}{\sqrt{2}} \sqrt{1+i\gamma} } \,\right) \,\right] \label{40}
\end{eqnarray}

Expression (\ref{40}) can be written in a compact form in terms of trigonometric functions as
\begin{equation}
V_{int}\left(\, r\,\right) 
=-\frac{e^2}{4\pi\, r} e^{-\frac{\mu_A r}{2} \gamma_+}\left[\,\cos\left(\, \frac{\mu_A r}{2} \gamma_-\,\right) 
+\frac{2\mu^2_B-\mu^2_A}{\mu_A^2\gamma_+\gamma_-} 
    \sin\left(\,\frac{\mu_A r}{2} \gamma_- \,\right)
 \,\right] \label{vfinal}
 \end{equation}

where we have introduced the notation

\begin{equation}
 \gamma_\pm \equiv \sqrt{\frac{2\mu_B}{\mu_A}\pm 1}\ ,\qquad \gamma =\gamma_+\gamma_-
\end{equation}

Now we consider the  limit where $\mu_A  << \mu_B$, which implies $\gamma_\pm\rightarrow \sqrt{2\mu_B/\mu_A}$. In this case,
(\ref{vfinal}) reduces to

\begin{equation}
 V_{int}\left(\, r\,\right)= -\frac{e^2}{4\pi r} +\frac{e^2 \mu^2_B}{8\pi} r 
-\frac{e^2}{4\pi}\frac{\mu_B^{3/2}}{\sqrt{2\mu_A}} +O\left(\,\mu_A \,\right)  \label{maslesslim}
 \end{equation}

One can notice that the last term in (\ref{maslesslim}) is infrared divergent for $\mu_A\to 0 $. 
In order to be able to consider the Cornell limit it is  necessary to regularize
$V_{int}\left(\, r\,\right)$ by making an appropriate choice of its zero. This is  perfectly legitimate procedure 
as the potential is always defined up to an arbitrary additive constant. 
Thus, we define the regular form of (\ref{vfinal}) as

\begin{eqnarray}
V_{reg}\left(\, r\,\right)&& \equiv -\frac{e^2}{4\pi\, r} e^{-\frac{\mu_A r}{2} \gamma_+}\left[\,\cos\left(\, \frac{\mu_A r}{2} 
\gamma_-\,\right) +\frac{2\mu^2_B-\mu^2_A}{\mu_A^2\gamma_+\gamma_-} 
    \sin\left(\,\frac{\mu_A r}{2} \gamma_- \,\right) \,\right]
 +\frac{e^2}{4\pi}\frac{\mu_B^{3/2}}{\sqrt{2\mu_A}}\nonumber\\
&& \label{44}
\end{eqnarray}

Now, we safely obtain the Cornell potential in the infrared limit of (\ref{44})

\begin{equation}
V_{reg}\left(\, r\,\right)\rightarrow -\frac{e^2}{4\pi\, r} +\frac{e^2\mu^2_B }{8\pi} r \label{C}
\end{equation}

The result (\ref{C}) clearly shows that the  Kalb-Ramond mass $\mu_B$ plays the role of ``~string tension~'' of the flux tube 
connecting the two opposite charges resulting in their linear confinement at large distances.\\
We mention that similar conclusions can also be obtained in a different ap-
proach based on the description of QED in terms of gauge invariant variables
\cite{Gaete:1998vr,Gaete:1999iy,Gaete:2007zn,Gaete:2007sj,Gaete:2009xf}.

If we allow the photon mass to be massive $0<\mu_A < 2\mu_B$, we notice that the long distance behavior of the potential
exhibit a horizontal asymptote for $r\to \infty$: 

\begin{equation}
 \lim_{r\to \infty} V_{reg}\left(\, r\,\right)=\frac{e^2}{4\pi}\frac{\mu_B^{3/2}}{\sqrt{2\mu_A}}
\end{equation}

Therefore, instead of permanent confinement for $\mu_A=0$, the two charges display  a spectrum of bound states 
up to a limiting energy given by the regularization constant. Above this energy the  pair can break up and charges are set free.
This behavior persists until  $\mu_A= 2\mu_B$, $\gamma_- = 0$, $\gamma_+ = \sqrt{2}$. 
At this point $ V_{reg}\left(\, r\,\right)$ becomes

\begin{equation}
V_{reg}\left(\, r\,\right)= -\frac{e^2}{4\pi\, r}   e^{-\frac{\mu_A r}{\sqrt{2} }}
\left(\, 1 - \frac{\mu_A r}{4\sqrt{2}}\,\right) +\frac{e^2}{16\pi}\, 
\mu_A \label{vlim}
\end{equation}

\begin{figure}[h]
\begin{center}
\includegraphics[width=10cm]{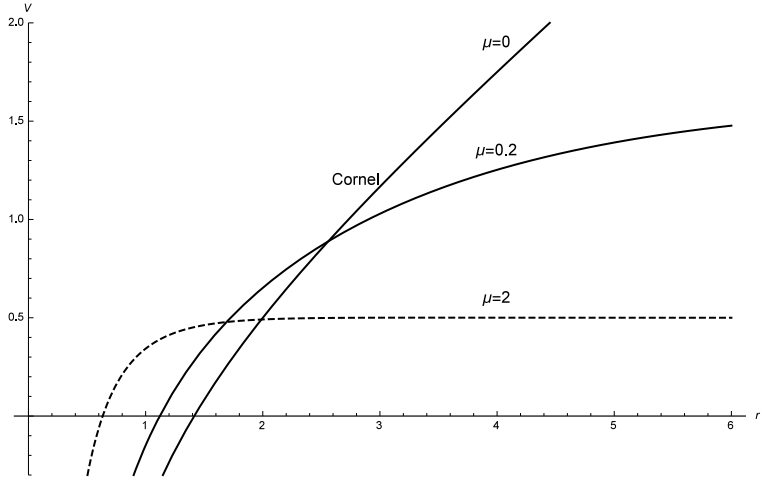}
\caption{ Plot of $V_{reg}\left(\, r\,\right)$ for different values of the ratio $\mu \equiv \mu_A/\mu_B$. 
$\mu=0$ is the Cornell
potential. The dotted curve represents the limiting case (\ref{vlim}).}
\label{CC}
\end{center}
\end{figure}

For transparency we have plotted  $V_{reg}\left(\, r\,\right)$  in Figure(\ref{CC}) in the range $ 0\le \mu_A\le 2\mu_B $.

\vspace{0.5cm}
\vspace{0.5cm}

It is interesting to see what happens if $\mu_A > 2\mu_B$. In this regime the poles (\ref{poli}) move to the real axis and
we have
\vspace{0.5cm}
\begin{figure}[h]
\begin{center}
\includegraphics[width=10cm]{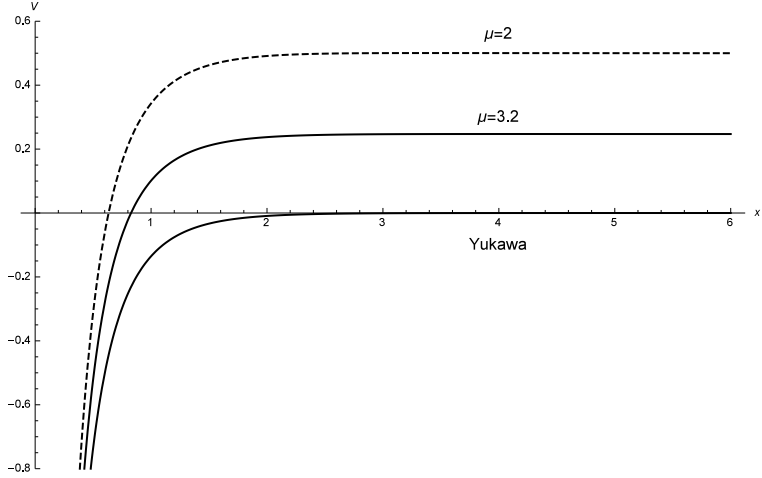}
\caption{Plot of the function in (\ref{v2}). $\mu \equiv \mu_A/\mu_B$, $\mu \ge 2$. }
\label{YYY}
\end{center}
\end{figure}
\vspace{0.5cm}

\begin{equation}
 \mu^2_A  > 4\mu^2_B\ , \quad m^2_\pm=\frac{\mu^2_A}{2}\left(\, 1 \pm \Gamma \,\right)\ ,\quad
\Gamma\equiv \sqrt{1 -\frac{4\mu^2_B}{\mu^2_A}}\equiv \Gamma_+ \Gamma_-
\end{equation}
 
with 

\begin{eqnarray}
&& \Gamma_\pm \equiv \sqrt{1\pm \frac{2\mu_B}{\mu_A}}\ ,\\
&& \gamma_+=\Gamma_+\ ,\\
&& \gamma_-=i\Gamma_-
\end{eqnarray}

In this case  the potential (\ref{44}) transforms into

\begin{equation}
V_{reg}\left(\, r\,\right) =-\frac{e^2}{4\pi\, r} e^{-\frac{\mu_A r}{2} \Gamma_+}
\left[\,  \cosh\left(\, \frac{\mu_A r}{2} \Gamma_-\,\right) +\frac{2\mu^2_B-\mu^2_A}{\mu_A^2\Gamma_+\Gamma_-} 
    \sinh\left(\,\frac{\mu_A r}{2} \Gamma_- \,\right)
 \,\right] +\frac{e^2}{4\pi}\frac{\mu_B^{3/2}}{\sqrt{2\mu_A}}
\label{v2}
\end{equation}

with the range for $\mu_B$ corresponding to  $0 \le \mu_B \le \mu_A/2$, for a given $\mu_A$.\\

\begin{itemize}

\item The first limit of interest for (\ref{v2}) is  $\mu_B\to 0 $, which leads to a Yukawa potential

\begin{equation}
V_{int}\left(\, r\,\right)\to
-\frac{e^2}{4\pi}\frac{1}{r}\, e^{-\mu_A r} \label{yuk}
\end{equation}

In addition, (\ref{yuk}) reduces to the Coulomb potential once we impose $\mu_A \to 0$.

\begin{equation}
V_{int}\left(\, r\,\right)= -\frac{e^2}{4\pi}\frac{1}{r}
\end{equation}

\item Going in the other direction by increasing $\mu_B \to \mu_A/2 $ in (\ref{v2}), it approaches (\ref{vlim}) 
from below as can be seen in Figure(\ref{YYY}).\\
This regime is characterized by the absence of linear confinement in favor of charges bound states up to the asymptotic energy
$E= e^2\mu_B/8\pi$. Above this energy the pair breaks and the charges are set free.
 
\end{itemize}

\section{Discussion and conclusions}
\label{ciacole}

In this paper we described an Abelian gauge theory exhibiting a confining potential at large distance. This unexpected
result, at least in ordinary Maxwell electrodynamics, is achieved by coupling the vector field to a Kalb-Ramond field.
Furthermore, both fields acquire mass via the Higgs mechanism  implemented through one complex and one neutral scalar field.
The Higgs fields are then frozen in their respective  non-trivial vacua in which dynamical 
fields propagate. Integrating out the Goldstone mode and the Kalb-Ramond field one obtains
a non-local effective theory for the  vector gauge field. 
The resulting model resembles Lee-Wick higher derivative electrodynamics 
\cite{bopp,Podolsky:1942zz,Podolsky:1944zz,Lee:1969fy,Lee:1970iw}, but with an important difference: the
former model was originally introduced in order to regularize short distance behavior of $QED$.  In the static limit.
the interaction potential energy is finite and linearly dependent on the distance between charges. The Coulomb behavior
is recovered at large distances.
On the other hand, confinement is characterized by the linear behavior but at large distances. \\
Phenomenologically, confinement is simulated by the the Cornell potential, which is a kind of
``~dual~''version of the Lee-Wick potential exchanging long range and short range regimes 
\cite{Smailagic:2001ch,Hjelmeland:1997eg}.
We were able to obtain this behavior providing that
the vector gauge field remains massless while the Kalb-Ramond field acquires mass through the Higgs mechanism. \\
The Cornell regime is part of a more general picture when both fields are massive. However, the massless photon 
limit has to be taken with caution since  the static potential contains  a constant term which is infrared divergent. 
To avoid this divergence one  exploits the arbitrariness in the definition of the zero of the  potential in order to define
 a finite infrared limit. 

\vspace{0.5cm}
\begin{figure}[h]
\begin{center}
\includegraphics[width=10cm]{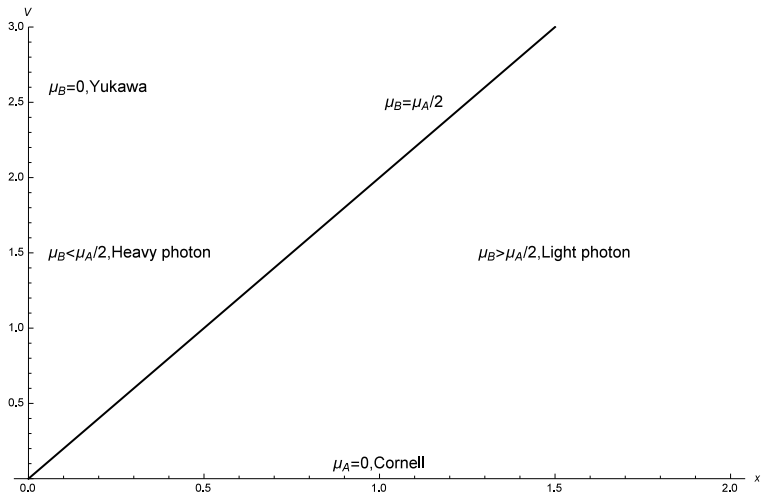}
\caption{Portrait of the vacuum phases in the model (\ref{l3}). The vertical axis corresponds to the Yukawa potential, 
along the horizontal axis is a linear potential. Finally, the origin represents the Coulomb phase. The inclined line marks
the boundary between the heavy and light photon phases.}
\label{Fasi}
\end{center}
\end{figure}
\vspace{0.5cm}

For the sake of clarity, we depicted in Figure(\ref{Fasi}) two different phases of the vacuum for the model described 
by the effective Lagrangian (\ref{l3}).
The order parameter is defined as $\mu\equiv \mu_A/\mu_B$. The horizontal axis $\mu_A=0$ describes the Cornell phase, 
while the vertical axis $\mu_B=0$ corresponds to the Yukawa phase. 
The two phases match along the transition line $\mu_A=2\mu_B$. The origin
of the coordinate system corresponds to the pure Coulomb limit. \\
 The choice of an Abelian model was made  in order to perform an explicit,
analytical, evaluation of the static potential energy between two electric charges. By itself this is a novel and
 interesting result. As it stands it may not be directly relevant for a realistic Yang-Mills $QCD$, but it may pave the way for
 a future investigation in that direction. This belief is additionally supported by several indications that, 
even in $QCD$, the confinement phenomenon is basically due to the Abelian sub-group of the complete non-Abelian 
symmetry group.


\end{document}